\documentclass[link]{IWCOMP}

\copyrightyear{2024}

\DOI{xxxxxx}

\begin{document}

\title[Advancing Similarity Search with GenAI]{Advancing Similarity Search with GenAI: A Retrieval Augmented Generation Approach}

\author{Jean Bertin}

\email{jean.bertin@mines-paris.org}

\shortauthors{Advancing Similarity Search with GenAI}

\begin{abstract}

This article introduces an innovative Retrieval Augmented Generation approach to similarity search. The proposed method uses a generative model to capture nuanced semantic information and retrieve similarity scores based on advanced context understanding. The study focuses on the BIOSSES dataset containing 100 pairs of sentences extracted from the biomedical domain, and introduces similarity search correlation results that outperform those previously attained on this dataset. Through an in-depth analysis of the model sensitivity, the research identifies optimal conditions leading to the highest similarity search accuracy: the results reveals high Pearson correlation scores, reaching specifically 0.905 at a temperature of 0.5 and a sample size of 20 examples provided in the prompt. The findings underscore the potential of generative models for semantic information retrieval and emphasize a promising research direction to similarity search.

\end{abstract}


\category{informatics; data science}

\maketitle

\section{Introduction}\label{sec1}\vspace*{5pt}

Similarity search is today widely employed across various applications, including document and image retrieval, product recommendation, sequencing search in bioinformatics or anomaly detection in cyber security. In general terms, similarity search technic consist in identifying and retrieving elements similar to a specific query within a data collection, using a metric measure to evaluate the proximity between those elements. The first similarity search methods were developed in the 1970s using inverted lists and have since significantly evolved. 

Original string-based methods, such as \textit{Levenhstein}, \textit{Q-gram}, \textit{Cosine}, \textit{Jaccard} work directly with the characters as a basis for the similarity metric. \textit{Jaccard} similarity assesses for instance the ratio between the intersection and union of two elements, while  \textit{Q-gram} divides a character sequence into fixed-length substrings to analyze their presence and frequency. In specific cases, these methods might be useful for comparing sequences of characters but encounter limitations when it comes to grasping the overall meaning of a sentence or paragraph.

The advent of vector databases marks a significant advance in representing data in the form of numerical vectors, enabling similarity search based on the meaning of elements rather than on exact matches. The key difference actually lies in data representation: string-based methods directly analyze characters, while vector-based methods transform data into continuous vectors, capturing semantic information. Vector-based approaches are now widely employed, and a number of frameworks like  \textit{Sent2Vec},  \textit{FAISS}, or  \textit{fastTextare} allow to create vector embeddings for sentences, capturing their meaning in a high-dimensional space. 

 \citealp*{bib1} investigated different similarity search techniques, including both string-based methods like \textit{Jaccard} or \textit{Q-gram}, and vector embedding methods, both supervised and unsupervised, combining multiple approaches. They surpassed previous similarity results on the BIOSSES dataset (\citealp*{bib2}) with a supervised combination of several methods (\textit{Jaccard}, \textit{Q-gram}, \textit{sent2vec}, \textit{Paragraph vector PV-DM}, \textit{skip-thoughts}, and \textit{fastText}). 

This article introduces an innovative similarity search approach through the utilization of a generative model to retrieve the distances between pairs of sentences. The article explore in fact how a \textit{Retrieval Augmented Generation} method (allowing to combine information retrieval with generative language modeling) can be used to create more accurate response in the context of similarity search. It should be noted that the presented results serve more as a starting point than an absolute performance assessment of this method, as not all generative models and prompt configurations could be tested : alongside the results, which are likely to evolve in the coming years with the improvements of Large Language Models (LLMs), this article primarily emphasizes the research direction and the innovative methodology proposed.

\section{METHOD}\label{sec2}

Most of recent similarity search techniques would traditionally use an \textit{embedding model} to create a vector database on the sentences to be compared: the similarity degree between these sentences is then retrieved using a metric to calculate the distances between vectors. As pointed out by \citealp*{bib5}, most of the string-based methods mentioned in the introduction, such as \textit{Jaccard}, \textit{Cosine}, or even a basic \textit{Euclidean} norm are in that case well suited to calculate distances between vectors. 

While \citealp*{bib4} already compared the performance of different embedding models on the BIOSSES dataset, this article focuses on examining results using the \textit{Retrieval Augmented Generation method}: instead of basing the similarity search on a distance computation method between vectors, the article propose to build a \textit{conversational chain} to evaluate each sentence pair similarity from the BIOSSES test dataset. The optimization of the prompt with the search for a perfect phrasing will therefore be equivalent to build the most optimized and personalized distance metric suited to the use case under study. Using the appropriate phrasing, the proposed \textit{Retrieval Augmented Generation} method enables consequently to spell out what distinguishes a small distance (i.e. the two phrases have very similar meanings) from a larger distance (i.e. the two sentences have more distant meanings).

\subsection{Presentation of the BIOSSES dataset}\label{subsec2.1}

The research will be conducted on the BIOSSES (Biomedical Semantic Similarity Estimation System) dataset created by \citealp*{bib2}. This dataset contains 100 pairs of sentences associated with a similarity score range from 0 to 4 (based on the guidelines of \textit{SemEval 2012 Task 6 on STS} proposed by \citealp*{bib3}) and evaluated by five different human expert. The sentences in the BIOSSES dataset are extracted from citations of biomedical articles in such a way that each pair of sentences either refers to the same or in the contrary to different articles : this design aims to cover a wide range of semantic similarities, from sentences closely related in meaning to sentences with more distant semantic connections. A short extract of two lines extracted from the BIOSSES dataset is presented below : 

\begin{table}[hbt!]
\centering
\begin{tabular}{p{3.2cm} p{3.2cm} S[table-format=1.1]}
\toprule
\textbf{Sentence 1} & \textbf{Sentence 2} & \multicolumn{1}{c}{\textbf{Score}} \\
\midrule
The oncogenic activity of mutant Kras appears dependent on functional Craf. & Oncogenic KRAS mutations are common in cancer. & 2.2 \\
\\
The up-regulation of miR-146a was also detected in cervical cancer tissues. & The expression of miR-146a has been found to be up-regulated in cervical cancer. & 4 \\
\bottomrule
\end{tabular}
\vspace{1em}
\caption{Extract of 2 lines from the train BIOSSES dataset.\label{tab1}}
\end{table}

In total, the BIOSSES dataset contains 100 lines, distributed as follows: 64 lines builds the train dataset, 16 lines the validation dataset and 20 lines the test dataset. 

\subsection{Calculation metric for similarity}\label{subsec2.2:pearson_correlation}

The accuracy of the similarity search method proposed in this article is evaluated by the \textit{Pearson correlation} between the similarity scores from the test dataset (also labeled \textit{reference similarity score}) and the similarity score provided by the generative model. In general terms, the \textit{Pearson correlation} between two variables $X$ and $Y$ can be calculated using the following formula:

\begin{equation}
r_{XY} = \frac{\sum_{i=1}^{n} (X_i - \bar{X})(Y_i - \bar{Y})}{\sqrt{\sum_{i=1}^{n} (X_i - \bar{X})^2 \sum_{i=1}^{n} (Y_i - \bar{Y})^2}}
\end{equation}

where:
\begin{itemize}
  \item $r_{XY}$ is the \textit{Pearson correlation} between $X$ and $Y$,
  \item $n$ is the number of observations,
  \item $X_i$ and $Y_i$ are the values of observations $i$ for $X$ and $Y$ respectively,
  \item $\bar{X}$ and $\bar{Y}$ are the means of $X$ and $Y$ respectively.
\end{itemize}

In this paper, the variables \textit{X} and \textit{Y} used to calculate the \textit{Pearson correlation}  are represented by two series of decimal numbers with values being contained between 0.0 and 4.0. Logically, the \textit{Person correlation} is equal to 1 when theses two series are strictly correlated, i.e. when the similarity scores retrieved by the model are proportional (or even equal in this research work, since the amplitudes are equal) to the \textit{reference similarity scores} furnished on the test dataset. 

\subsection{Prompt engineering for similarity search case}\label{subsec2.3:prompt_engineering}

The usage of a generative model to retrieve similarity scores between pairs of sentences from the BIOSSES dataset requires a suitable prompt, i.e. a set of clear instructions enabling the model to perform the task as accurately as possible. \citealp*{bib9} investigated the effects of six prompt engineering techniques to generate ideas for an exemplary scenario, highlighting the impact of the used prompting technique to retrieve the expected result. In the context of this study, the prompt is actually divided into two categories : the \textbf{\textit{system prompt}} and the \textbf{\textit{user prompt}}. 

The \textbf{\textit{system prompt}} corresponds to the general instructions: it gives the context for the execution of the \textit{conversational chain} build from the generative model. In the context of this research, the \textit{system prompt} also includes training examples extracted from the train dataset to give the model a better understanding of how similarity scores are determined. The \textit{gpt-3.5-turbo} model has been used, but as mentioned by \citealp*{bib12}, many other commercial (\textit{Bard, Claude, Jurassic}, etc.) or open-source (\textit{Alpaca, BLOOM, Dolly, Llama 2}, etc.) generative models are already available and could have been used for this study.\footnote{The LLM Index : list of large language models, including open-source and commercial offerings(https://sapling.ai/llm/index)} Comparing the result accuracy across these different generative models is not the purpose of this article but could be a subject of future researches.

The \textit{system prompt} is defined as follows:

\begin{center}  
    \par 
    \textit{You are a helpful assistant who helps retrieve similarity scores between two sentences.
You will find below some examples to help you determine this similarity score with the best accuracy:
 ...}
\end{center}

The proposed method allows then to add directly after this prompt a set of examples extracted from the training dataset to be included in the \textit{system prompt}. It is actually not always relevant to integrate the largest possible number of examples since the model has already been trained over billions of parameters, and there is a risk of confusion and misinterpretation when providing an overly extensive prompt: a sensitivity study of the model in relation to this criterion is therefore proposed in Section \ref{sec:results}. \nameref{sec:results}. Since the generative models are already trained on billions of parameters and may also provide accurate responses even without any \textit{fine-tuning}, the article also include a similarity analysis without any training examples being injected on the \textit{system prompt}.

The \textbf{\textit{user prompt}} allows then to provide the model with a specific problem-solving scenario using the test dataset. The \textit{user prompt} is defined as follows:

\begin{center}
    
    \par 

    \textit{Please give me the similarity score from 0 to 4 between those sentences: \textbf{Sentence1} and \textbf{Sentence2}. \\
        Always respond using strictly and only the following format: Similarity score : ...}
\end{center}

\textit{\textbf{Sentence1}} and \textit{\textbf{Sentence2}} corresponds to the sentence pairs for which the similarity score must be retrieved. The second phrase of the \textit{user prompt} is crucial as it defines the expected output format of the response provided by the generative model. Since the similarity score can be further processed only if the right formalism has been delivered, it is important to restrict the model's freedom regarding the expected form of results. The words \textit{always},\textit{strictly}, and \textit{only} are precisely employed on that purpose.

\subsection{Implementation of the conversational chain}\label{subsec2.4}

The \textit{conversational chain} can easily be set up using a \textit{completion chat function} taking as arguments the \textit{system prompt} and the \textit{user prompt} defined in the previous paragraph, and as well the generative model for which the \textit{temperature} parameter must be set up. The impact of this \textit{temperature} parameter on the result accuracy is analysed in detail in the paragraph \ref{subsec3.1:temperature_sensibility}.

The answer provided by the generative model will then be processed using regular expression operators (\textit{regex}) and converted to decimal format in order to be integrated as a decimal value inside a new column entitled \textit{model\_score}. The output dataset then contains the two original columns \textit{Sentence1} and \textit{Sentence2} from the test dataset, and as well as two similarity score columns, the first one corresponding to the \textit{reference similarity score} and the second one being provided by the generative model.

\subsection{Iterate on test dataset}\label{subsec2.5}

An iteration is then carried out on each sentence pairs from the test dataset, using the \textit{conversational chain} described in the previous paragraph to determine a similarity score for each of them. A new \textit{conversational chain} is therefore created at each iteration, the only difference at each iteration concerning the \textit{user prompt} being updated with a new sentence pairs at each iteration.
A formalism test is also implemented to ensure that all responses retrieved by the model corresponds to the expected output : the iteration will simply move on to the next sentence pairs if the formatting conditions of the response are respected.

A schematic diagram of the iteration purpose is represented below: 

\begin{figure}[ht!]
  \centering
  \includegraphics[width=1.0\linewidth]{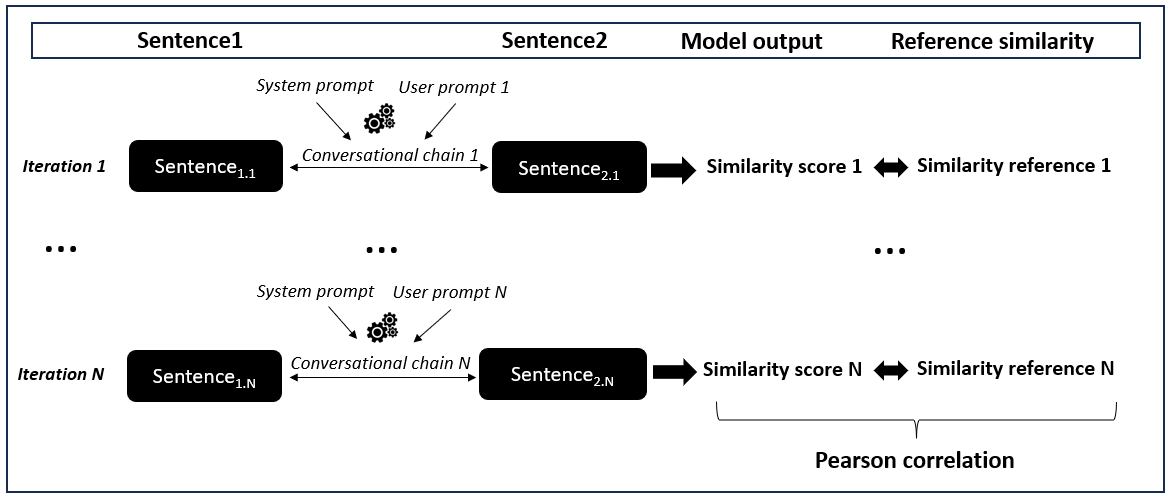} 
  \caption{Conversational chain iteration on pairs of sentences\label{fig1}}
  \vspace*{-3pt}
\end{figure}

As illustrated in Figure \ref{fig1}, the calculation of the \textit{Pearson correlation} specified in paragraph \ref{subsec2.2:pearson_correlation} occurs at the end of the iteration loop, i.e. when all sentence pairs from the test dataset have been passed through, and the generative model has provided a similarity score for each of them.

\section{RESULTS}\label{sec:results}

\textbf{The first analysis focuses on the sensitivity of the results to the temperature parameter, whereas the second part analyzes the model's sensitivity to the number of examples given to the system prompt}.

\subsection{Sensibility to temperature parameter}\label{subsec3.1:temperature_sensibility}

\begin{figure}[ht!]
  \centering
  \includegraphics[width=1.0\linewidth]{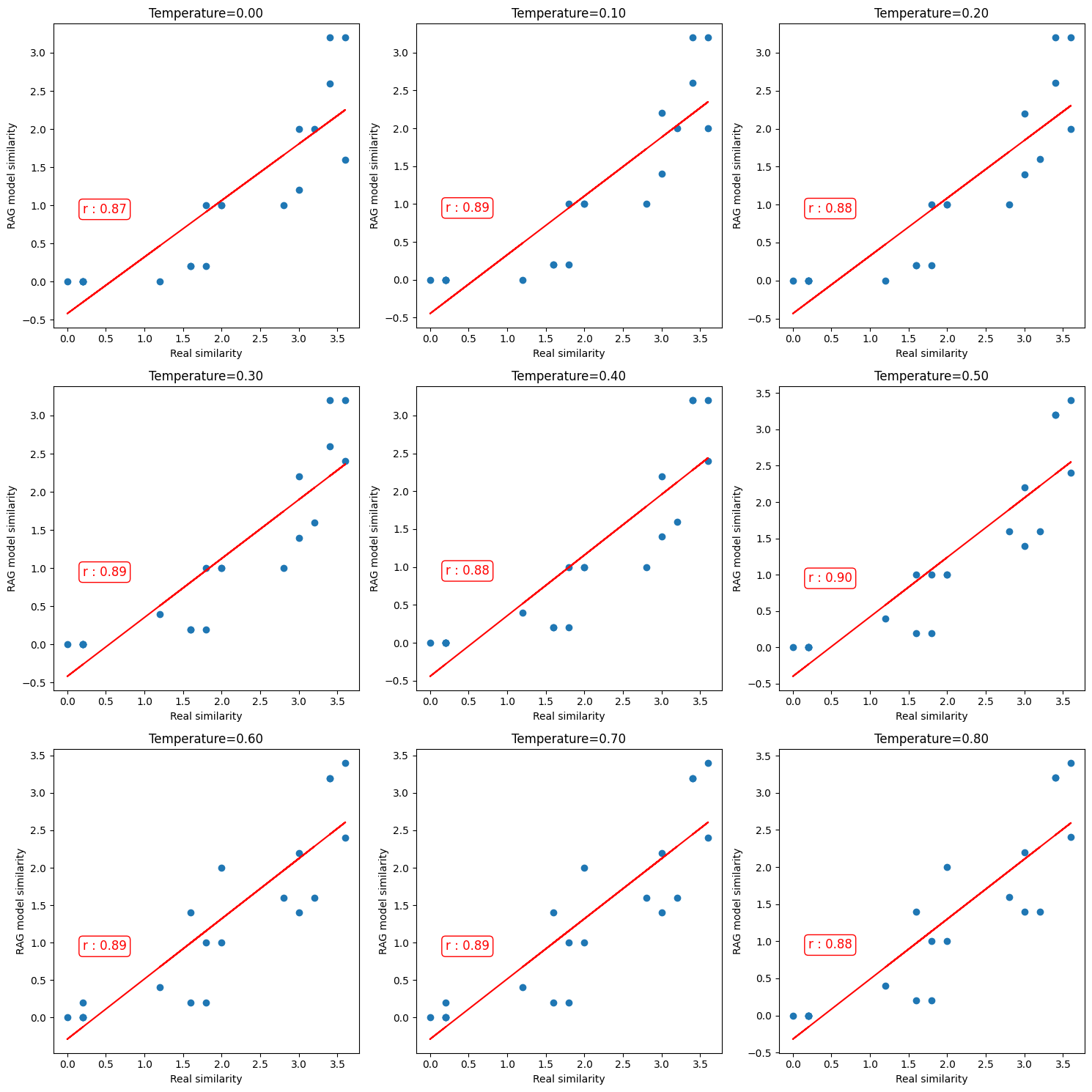} 
  \caption{Evolution of similarity results with temperature\label{fig2}}
  \vspace*{-3pt}
\end{figure}

For this study, an iteration is implemented to test different \textit{temperature} values and to calculate the corresponding similarity score obtained on each line of the test dataset. The \textit{Pearson coefficient} is calculated for each temperature value from 0 to 1 with an step of 0.1, and the associated graph is plotted (see figure \ref{fig2}).

The highest \textit{Pearson coefficient} is equal to \textbf{0.905} and is obtained for a \textit{temperature} value of 0.5. This \textit{temperature} value introduces some diversity into the predictions, which is beneficial for similarity calculations to take into account minor variations between sentences. This \textit{temperature} of 0.5 also give the model some flexibility to take into account the language nuances and the implicit meanings that might be related to certain phrasing constructs. 

The risk to use high \textit{temperature} values, however, concerns the risk of overstepping the formalism of the response provided by the model, underlining the importance of the choice of terms used in the prompt as explained in paragraph \ref{subsec2.3:prompt_engineering}.

\subsection{Influence of the number of examples given to the prompt}\label{subsec3.2}

The results now focuses on the sensitivity of the model concerning the number of examples provided in the \textit{system prompt}. These examples are extracted from the train dataset and added next to the \textit{user prompt} using the following formalism:

\begin{center}  
    \par 
    \textit{The sentence "The oncogenic activity of mutant Kras appears dependent on functional Craf." and the sentence "Oncogenic KRAS mutations are common in cancer" have a similarty score of \textbf{2.2}}
\end{center}

Similar to the \textit{temperature} parameter evaluation, an iteration is constructed over different sample sizes of examples ingested in the \textit{system prompt}, allowing the measurement of the \textit{Pearson coefficient}. A scatter plot for each sample size of examples from 0 to 60 with a step of 10 is then plotted to compare the \textit{reference similarity score} to the similarity obtained by the model.

\begin{figure}[ht!]
  \centering
  \includegraphics[width=1.0\linewidth, height=0.25\textheight]{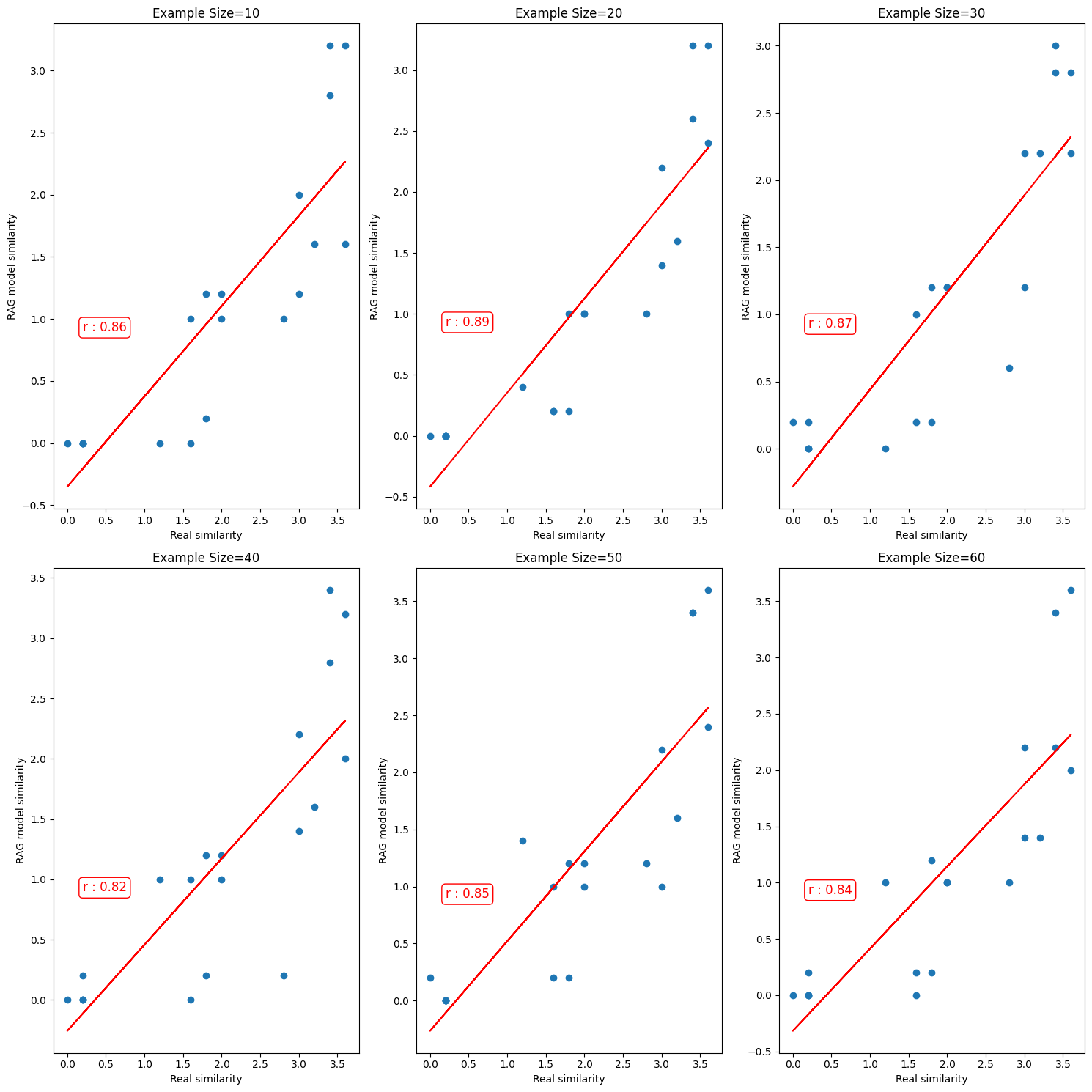} 
  \caption{Evolution of similarity results with the number of examples given to the prompt\label{fig3}}
  \vspace*{-3pt}
\end{figure}

The highest \textit{Pearson correlation} is equal to \textbf{0.89} and is obtained for a sample size in the \textit{system prompt} of 20 examples. As showed by \citealp*{bib7}, the inclusion of examples within the prompt improves the response accuracy by helping the model to precisely understand the expected task. This consideration is related to the \textit{adversarial prompt} research field which consist to provide \textit{adversarial} examples to the model (i.e. slight modifications of correctly classified inputs) to avoid miss-classification when the response is given (\citealp*{bib8}).

\subsection{Cross-factor analysis}\label{subsec3.3}

In order to observe the model sensitivity related to the combined variation of the two factors under study (\textit{temperature} and number of examples supplied in the \textit{system prompt}), a table providing \textit{Pearson correlation} values has been retrieved: \textit{temperature} values are still incremented from 0 to 1 with a step of 0.1, while sample size of examples supplied to the prompt system is incremented from 0 to 60 with a step of 10.

\begin{figure}[ht!]
  \centering
  \includegraphics[width=1.0\linewidth]{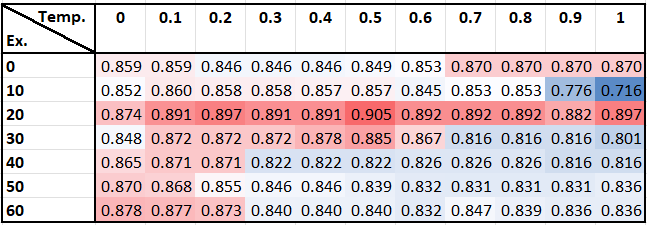} 
  \caption{Pearson correlation as function of temperature and sample size\label{fig4}}
  \vspace*{-3pt}
\end{figure}

As shown on Figure \ref{fig4}, the highest \textit{Pearson correlation} is equal to \textbf{0.905} and is obtained for a temperature value of 0.5 and a sample size of 20 examples. This result appears to exceed the correlation value of \textbf{0.871} obtained by the \textit{supervised combination of several methods} proposed by \citealp*{bib1} and consequently, also surpasses the correlation results obtained earlier by \citealp*{bib4} and \citealp*{bib2}.

\section{DISCUSSION}\label{sec5}

\subsection{Limitations and constraints}\label{subsec5.1}

An apparent limitation of the proposed architecture concern the computational resources required to process the data for in an iterative way: each iteration requires indeed the \textit{conversational chain} to be rebuilt on the basis of the adapted \textit{user prompt}. As a result, this solution is neither optimized in terms of computational resources, nor in terms of financial cost as soon as the utilization of the generative model may be subject to a charge. This iteration can however not be avoided due to the limited number of tokens (related to the \textit{state of the art} generative models) which restrict the size of the answer. In addition, it would be difficult to make the model understand on a \textit{one-shot} request (i.e. without iteration architecture) the necessity to retrieve the totality of the similarity scores at once.

Another limitation concerns the sensitivity of the model to the expected output for the similarity score. Even with the \textit{temperature} parameter being set to 0 and despite the fact that the \textit{Retrieval-Augmented Generation} is a promising approach for mitigating the large language models \textit{hallucination} (\citealp*{bib14}), there's no strict guarantee to avoid any \textit{noise} or \textit{counterfactual robustness} that would conduct the generative model to ignore the requested output format, or to give a score based on criteria that would not have been defined in the prompt. 

The last limitation concern the non-reproducibility of the results. Even if it has never occurred during the tests for this research (particularly due to the use of the \textit{seed} command in the build of the conversational chain), the non-deterministic nature of LLMs models implies that the same input prompt might produce different responses over different runs. The evaluation of the method would therefore be easier with a methodology enabling an absolute guarantee to get the results being fixed and no longer influenced by model's internal state or specific conditions under which it operates. 

\subsection{Areas for improvement}\label{subsec5.2}

Many improvements for the presented similarity search method could be proposed, starting with the optimization of the used prompt to retrieve the similarity. \citealp*{bib6} research demonstrated how the use of prompt engineering techniques such as \textit{role-prompting}, \textit{one-shot}, and \textit{few-shot prompting}, and even more sophisticated practices such as \textit{chain-of-thought}, can improve the overall performance of LLMs. 

Another area for improvement concerns the type of generative model used. As mentioned in paragraph \ref{subsec2.3:prompt_engineering}, a vast catalog of generative models more or less adapted to the similarity search case already exist.  Through simulations and human experiments, \citealp*{bib10} showed for instance that the \textit{BART-Gen} model produced more human-like responses for generative inference than BERT, a popular model for natural language processing. These results demonstrate just how essential explicit representations are in human generative reasoning.

A last area for improvement concerns the extension of the proposed similarity search method to other study variants : as proposed by \citealp*{bib11}, the advanced semantic understanding of generative models could by used to address the \textit{nearest search case}. The proposed method in this article could therefore draw inspiration to address the case of nearest neighbor search using a generative model instead of traditional methods such as \textit{K-Nearest Neighbor algorithm}. In the context of the studied case, it would be necessary to convert the entirety of the second column (\textit{Sentence 2}) into a vector database. Then, an iteration would be performed on each element of the first column (\textit{Sentence1}), asking the model which of the elements stored in the vector database would have the closest meaning. In an iterative way, this method would therefore allow to find for each entity from the first table the closest related element of the second table.

\section{CONCLUSION}\label{sec6}

This article has introduced an innovative approach to address the similarity search case using a generative model. The \textit{Retrieval Augmented Generation} method has been explored to construct a \textit{conversational chain}, enabling an advanced understanding of sentences and consequently providing accurate similarity scores. 
An in-depth analysis was conducted to assess the model's sensitivity to parameters such as temperature and the number of examples included in the system prompt. The results demonstrated that moderate temperature values and an optimal number of examples in the system prompt lead to high Pearson correlation scores, surpassing the performance of existing similarity search methods.
However, some limitations were identified, particularly in terms of computational resources and the model's sensitivity to the expected output. Improvement suggestions were proposed, including the optimization of the prompt used and the exploration of more advanced prompt engineering techniques.

In conclusion, this research represents a significant advancement in the field of similarity search by leveraging the capabilities of generative models. The promising results offer potential development for future applications in any domain requiring a nuanced understanding of semantic similarities.

\makeatletter
\def\@biblabel#1{}
\makeatother

\section*{NOTE}
The Jupyter Notebook containing the python code (and the BIOSSES dataset) implemented for this paper is available on Github.\footnote{link: https://github.com/JeanBertinR/Similarity\_search\_article}

\end{document}